# ON A PHYSICAL METATHEORY OF CONSCIOUSNESS

Miroljub Dugić[1,4], Dejan Raković[2,4], and Milan M. Ćirković[3,5]

[1]Department of Physics, Faculty of Science, P.O.Box 60, 34000 Kragujevac, Yugoslavia
E-mail: dugic@knez.uis.kg.ac.yu ; Fax: (+381 34) 335-040

[2]Faculty of Electrical Engineering, P.O.Box 35-54, 11120 Beograd, Yugoslavia
E-mail: rakovic@net.yu ; Fax: (+381 11) 324-8681

[3]Astronomical Observatory, Volgina 7, 11000 Beograd, Yugoslavia
E-mail: arioch@eunet.yu ; Fax: (+381 11) 419-553

[4]International Anti-Stress Center, Smiljanićeva 11/III/7, 11000 Beograd, Yugoslavia
E-mail: info@iasc-bg.org.yu ; Web site: www.iasc-bg.org.yu; Fax: (+381 11) 444-7646

[5]Department of Physics and Astronomy, SUNY at Stony Brook,
Stony Brook, NY 11794-3800, USA

**Abstract.** We show that the modern quantum mechanics, and particularly the theory of decoherence, allows for formulating a sort of a physical metatheory of consciousness. Particularly, the analysis of the necessary conditions for the occurrence of decoherence, along with the hypothesis that consciousness bears (more-or-less) well definable physical origin, leads to a wider physical picture naturally involving consciousness. This can be considered as a sort of a psycho-physical parallelism, but on rather wide scales bearing some cosmological relevance.

## 1. INTRODUCTION

In this paper we want to point out that modern quantum mechanics allows for formulating a physical metatheory (metaphysical theory) of consciousness. This observation comes only from some recent progress in the foundations of the so-called decoherence theory [1-3]. In addition, this program is important in view of the contemporary heated debate [4] of reductionism versus holism in the philosophy of science.

We employ practically universally accepted hypothesis in physical considerations devoted to the issue of consciousness: there is a physical background (and/or physical basis) of consciousness which, as a physical system, can be described and treated by the methods of the physical sciences. This, partially trivial assertion will later on prove useful for our considerations, finally leading to a wider physical picture naturally involving consciousness, and eventually pointing out something new as regards the connection between physics and (the physics of) consciousness. As will become clear below, this reductionist attitude is justified exactly because quantum mechanics (which we use as a physical basis for discussion) is generally percieved as introducing a substantial holistic element of modern physics. Therefore, by pointing out elements necessary for building a metatheory of consciousness, we may bridge the gap between these two positions, as well as explore the limits of theory making process [5].

There is of course no big practical use of the metatheories, generally speaking. But the observations this way provided usually enrich and/or widen our point(s) of view. In our opinion, probably the main point of the present paper is that such a theory - metaphysical theory of consciousness - naturally follows from the foundations of quantum mechanics.

## 2. A BRIEF ACCOUNT OF THE THEORY OF DECOHERENCE

The most general physical situation in the decoherence theory [1-3,6,7] is as follows: There is a (quantum) system (S) which is in unavoidable interaction with its environment (E). The system S is so an open quantum system to which the unitary (reversible, Schrodinger) evolution in time cannot be ascribed. For the composite system "system + envirnment (S+E)", one applies the hypothesis of the universal validity of quantum mechanics, i.e. that the system S+E evolves according to the Schrodinger law.

That is, the unitary evolution in time of the composite system S+E is generated by the Hamiltonian:

$$\hat{H} = \hat{H}_S + \hat{H}_E + \hat{H}_{int} \tag{1}$$

where $\hat{H}_i$, $i = S, E$ represent the "self-Hamiltonians" of mutually noninteracting systems, and the term $\hat{H}_{int}$ represents their interaction (the interaction Hamiltonian). For the conservative systems, the corresponding unitary operator of evolution in time for the composite system is given by the expression:

$$\hat{U}(t) = \exp(-2\pi i \hat{H}/h). \tag{2}$$

Usually, for some plausible physical reasons and for mathematical simplicity (but without loss of generality), one adopts the following simplification:

$$\hat{U}(t) \cong \hat{U}_{int}(t) = \exp(-2\pi i \hat{H}_{int}/h). \tag{3}$$

Now, the main task of the decoherence theory is to calculate the subsystem's (S's) "density matrix", $\hat{\rho}_S$ defined by:

$$\hat{\rho}_S = tr_E \left( \hat{U}_{int}(t) |\Psi(t=0)\rangle_{SE\,SE}\langle\Psi(t=0)| \hat{U}^*_{int} \right) \tag{4}$$

where: "$tr_E$" denotes "tracing out" (i.e. integrating over) the environmental degrees of freedom, $|\Psi(t=0)\rangle_{SE}$ represents the initial state of the composite system, and "*" denotes the hermitian conjugate.

The "symptom" of the occurrence of decoherence (i.e. of the decoherence effect) is that there exists an orthonormalized basis $\{|i\rangle_S\}$ in the Hilbert state space of the system S, for which one obtains at least approximately diagonal form of the density matrix; such a basis is sometimes [6] referred to as to the "pointer basis" of the system S. The diagonalization is mathematically presented by:

$$\rho_{Sii'} \cong 0, i \neq i' \tag{5}$$

where $\rho_{Sii'} \equiv {}_S\langle i'|\hat{\rho}_S|i\rangle_S$.

Most of the operational tasks in the decoherence theory are of the following kind: to find out the "pointer basis" for the system S; i.e. to find the set of mutually at least approximately orthogonal states for the system S to which eq.(5) applies.

However, it has been shown recently [1-3] that the occurrence of decoherence substantially depends on some yet general characteristics of the interaction Hamiltonian. Particularly, the existence of the necessary conditions for the occurrence of decoherence has been pointed out. Physically, this result represents the necessary conditions for the existence of the "pointer basis" of the open system S.

## 3. IS NONEXISTENCE OF THE "POINTER BASIS" PHYSICALLY RELEVANT?

Nonexistence of the "pointer basis" - which mathematicaly follows from the quantum mechanical formalism[1] - can be interpreted[2] as that one cannot put a definite border-line between the two subsystems, S and E, of the composite system S+E. In the "macroscopic context" [7-10], this can be considered unphysical (or counterfactual), so bearing no physical significance, at least as regards the "realistic" physical systems (models).

However, this is not all that can be told in the "macroscopic context". To show this, we will briefly overview below the results of Dugić [1-3].

Actually, if one cannot define the border-line between[3] S and E, one may think of the cannonical transformations in the composite system S+E as a whole, so eventually obtaining the following physical situation: there exists a pair of the systems, S' and E', whose mutual interaction allows for the occurrence of decoherence, i.e. for defining the "pointer basis" of the new system S', but only simultaneously with defining the new environment E'. More precisely, we have the situation presented by the following scheme:

$$\{\textit{improper interaction in } S+E\} \xrightarrow{canonical-transformations} \{\textit{proper interaction in } S'+E'\} \qquad (6)$$

where "proper" ("improper") interaction means that one may (may not) define the "pointer basis" of the system of interest, S' (S).

It is crucial to note that both the new system S' and its (new) environment E' are - if at all - only simultaneously defined. As to the "old" systems, S and E, one may say that they stay undefined (plausibly[4]: unobservable). It is also crucial to note that the canonical transformations in eq.(6) are such that the degrees of freedom[5] of both S' and E' are (analytical) functions of the degrees of freedom of both S and E. Mathematically, this reads, e.g.:

$$\xi_{S'i} = f_i(x_{S\alpha}, p_{S\alpha}; X_{E\beta}, P_{E\beta})$$
$$\pi_{E'j} = f_j(x_{S\alpha}, p_{S\alpha}; X_{E\beta}, P_{E\beta}) \qquad (7)$$

with obvious notation.

The inverse transformations are physically irrelevant for the two reasons. First, these are meaningless due to nonexistence of the border-line in the system S+E. Second, knowing the value of, e.g., the coordinate $x_{Si}$, does not follow from knowing the values of the degrees of freedom of the system S'+E' - which is due to incompatibility of the observables of S'+E'. Therefore, the transformation eq.(6) is substantial: the (sub)systems S' and E' are "objective", while the "old" systems S and E are unobservable (cf. footnote 4).

---

[1] Once more speaking of the fact that, usually, quantum mechanics offers us more than we can classically understand, or expect.
[2] See Zurek [6], and for more detailed discussion see Dugić [3].
[3] I.e. if one cannot define the subsystems S, and E, through their (a priori given) degrees of freedom.
[4] Cf. Zurek [6].
[5] Coordinates and momenta.

## 4. THE ROOTS OF THE PHYSICAL METATHEORY OF CONSCIOUSNESS

When the system S' is a macroscopic (many-particle) system, the results of the previous section strongly support [3] the following statement: no a local action in the system S'+E' could help in "objectification" (i.e. in physical "appearance") of the subsystems S and E.

This notion follows from the fact that, as assumed, S'+E' is not an isolated[6] system. Then the interactions of S' and E' with the surrounding physical systems lead to a holistic nature of the complete system (which can be considered as isolated): a local interaction in the system S'+E' determines analogous interaction in a distant place[7]. The isolated system can be referred to as the macroscopic part of the Universe (MPU). Then, the MPU is interconnected so that no a local action can change the definition of its parts. But what about the global actions - i.e., of the global changes (the transformations inverse to eq.(6))? Interestingly, the answer is: such global transitions are unobservable.

To illustrate this, let us employ the hypothesis made explicit in Section 1. The physical basis of consciousness - which is a necessary part of the act of the "observation" - is also a macroscopic system. The global transition inverse to eq.(6), by definition, involves the degrees of freedom of this system, for it is considered [11,12] as an open system. Therefore, the global transitions lead to redefining of the physical basis of consciousness, thus giving rise to the following, a wider physical picture: each "Universe" defined by its parts, S+E, or S'+E', be it "objective" or not, defines a corresponding "kind" of consciousness - through definition of a physical system which is assumed (cf. Section 1) to be its physical basis. But this gives the roots of the physical metatheory of consciousness: consciousness is a relative concept, not independent on the (quite general) definition of the MPU. I.e., the physical basis of consciousness can be defined only simultaneously with the "rest" of the "Universe", in accordance with the holistic nature of the MPU as pointed above. This is not so surprising, especially if one takes seriously numerous anthropic "coincidences" playing a role in both classical and quantum cosmology [13]. This, in our opinion, is another feature of the "psycho-physical parallelism", and a sort of the relative-physical-theory of consciousness, which defines consciousness only in relation to the definition of the macroscopic part of the Universe.

As a consequence, one could conjecture that consciousness might be the essential property of Nature at different structural levels (macroscopic and microscopic, animate and inanimate), as widely claimed in traditional esoteric knowledge [14] - which might be supported by analogous mathematical formalisms of the dynamics of Hopfield's associated neural networks and Feynman's propagator version of Quantum mechanics [15]. Such nonlocal pantheistic idea of consciousness is also supported by Raković´s physical model of altered and transitional states of consciousness, which might provide additional route to the physical solution of the problem of the wave-packet reduction in the Quantum measurement theory too [11,16,17].

---

[6] This is not in contradiction with the expression (2). Actually, the interaction of a "system" with its environment dominates so giving rise to fast decoherence effect, while all the other interactions can be neglected during the time intervals of the order of the "decoherence time". This perturbation-like situation is a general feature of the decoherence theory: the open systems suffer very fast decoherentization, while the remaining dynamics is driven by their mutual interactions - as particularly stressed in Section 4 of Ref. [10].

[7] Interaction of S'+E' determines interaction, but also the definition of a remote system S''+E'', and vice versa.

## 5. CONCLUSION

It is pointed out that recent progress in decoherence theory implies that the physical basis of consciousness can be defined only simultaneously with the "rest" of the "Universe", in accordance with the holistic nature of the MPU as pointed above. This, in our opinion, is a feature of the "psycho-physical parallelism", and a sort of the relative-physical-theory of consciousness, which defines consciousness only in relation to the definition of the macroscopic part of the Universe.